\documentclass[12pt,preprint]{aastex}
\shorttitle{The Blue Tilt}
\shortauthors{Strader \& Smith}
\def\etal{{\it et al.}}

\begin{document}

\title{On The Origin of the Blue Tilt in Extragalactic Globular Cluster Systems}

\author{Jay Strader\altaffilmark{1,2}, Graeme H. Smith\altaffilmark{3}}
\email{jstrader@cfa.harvard.edu, graeme@ucolick.org}

\altaffiltext{1}{Harvard-Smithsonian Center for Astrophysics, Cambridge, MA 02138}
\altaffiltext{2}{Hubble Fellow}
\altaffiltext{3}{UCO/Lick Observatory, University of California, Santa Cruz, CA 95064}

\begin{abstract}

Some early-type galaxies show a correlation between color and integrated magnitude among the brighter 
metal-poor globular clusters (GCs). This phenomenon, known as the blue tilt, implies a mass-metallicity 
relationship among these clusters. In this paper we show that self-enrichment in GCs can explain several 
aspects of the blue tilt, and discuss predictions of this scenario.

\end{abstract}


\section{Introduction}

Several recent photometric studies of extragalactic globular cluster (GC) systems have discovered a novel 
feature. The blue, metal-poor population of GCs shows a color-magnitude relation at bright magnitudes: the 
more luminous GCs are redder. This ``blue tilt" was first noticed in the GC systems of massive elliptical 
(E) galaxies (Harris \etal~2006; Strader \etal~2006), but has since been seen in the Sa/S0 NGC 4594 
(Spitler \etal~2006), fainter early-type galaxies in Virgo (Mieske \etal~2006), and even in normal spirals 
(Strader \etal~2008, in preparation). Figure 1 shows the blue tilt in M87 (Strader \etal~2006; see also 
Mieske \etal~2006). The blue tilt is common but not ubiquitous---the massive E NGC 4472 does not have it, 
for example, and among galaxies that do show a tilt, the slope varies. There is no evidence for a tilt in 
the red, metal-rich subpopulation of GCs.

Many explanations for this phenomenon have been proposed. These include (i) self-enrichment, (ii) 
``pollution" of the GC sequence with objects such as dwarf galaxy nuclei or merged star clusters (e.g., 
Bekki \etal~2007), and (iii) gravitational capture of metal-rich field stars by GCs (Mieske \etal~2006). A 
number of other scenarios are discussed in Mieske \etal~(2006) but are thought by them to be much less 
probable.

Option (ii) is considered by both Mieske \etal~(2006) and Strader \etal~(2006), who conclude that it is 
unlikely. While the nuclei of Virgo dwarfs show their own color-magnitude relation (Lotz \etal~2001), the 
red mean colors of the luminous GCs are not due solely to the addition of bright red objects, but instead 
come from systematic shifts in the GC color distribution with magnitude. In addition, the number of nuclei 
needed is much larger than the handful of stripped nuclei that have been found in either the Virgo or 
Fornax clusters (see also discussion in \S 4).

Except for a few objects, the sizes of the blue tilt GCs are generally small, inconsistent with the 
expectations of merged clusters. Merged clusters should be larger as the orbital energy of the binary 
cluster is converted into kinetic energy in the merger product. The best candidate for a merged cluster, 
NGC 1846 in the Large Magellanic Cloud (Mackey \& Broby Nielsen 2007), has an unusually large projected 
half-light radius of $\sim 15$ pc (D.~Mackey, private communication). While it seems likely that such 
exotic objects contribute to the GC systems of massive galaxies at some level, especially at the brightest 
magnitudes (Harris \etal~2006; Strader \etal~2006), they are unlikely to be primarily responsible for the 
blue tilt.

Option (iii) was investigated using collisional N-body simulations by Mieske \& Baumgardt (2007). They 
report that field star capture is inefficient even in optimistic circumstances, and conclude that this is 
not the cause of the blue tilt.

Our working assumption is that self-enrichment remains the most likely explanation for the blue tilt. In 
this case the basic explanation is that the color-magnitude relation reflects a physical mass-metallicity 
relation for metal-poor GCs. As discussed in Strader \etal~(2006), there are a variety of GC 
self-enrichment models and simulations in the literature (e.g., Morgan \& Lake 1989; Brown \etal~1991; 
Parmentier \etal~1999; Parmentier \& Gilmore 2001; Parmentier 2004; Recchi \& Danziger 2005). Here we 
refer to models that focus on self-enrichment as the primary determinant of cluster metallicity, and not 
to the larger body of work on the origin of abundance anomalies in Galactic GCs that primarily involve 
light elements such as C, N, O, and Mg. Among GC self-enrichment models, there is no consensus on the 
dominant physical processes or expected level of enrichment as a function of cluster mass. The lack of a 
tilt in the red GCs might be expected if the self-enriched metals were a constant fraction of the stellar 
mass, since the overall metal content of red GCs is a factor of $\sim 10$ higher than the blue GCs 
(Strader \etal~2006; Mieske \etal~2006). We illustrate this point further in \S 3.

A possible relationship between GC self-enrichment and a mass-metallicity relation among GCs was discussed 
for the oldest halo Galactic GCs by Parmentier \& Gilmore (2001), prior to the discovery of the blue tilt 
in external galaxies. However, Parmentier \& Gilmore were attempting to explain an anticorrelation between 
GC metallicity and mass, the opposite of the blue tilt.

Here we do not attempt detailed modeling of the self-enrichment process (see the studies cited in the 
previous paragraphs for such models). Our model differs from many of those previously cited in that we 
invoke a single generation of star formation; within the context of self-enrichment, we expect a spread 
in metallicity in individual star clusters. This is inconsistent with the chemical homogeneity observed 
in most Galactic GCs, but the Galaxy does not have a blue tilt (see additional discussion in \S 8).

Our approach is to use scaling relation assumptions about the putative enrichment to explore the 
conditions under which the observed GC mass-metallicity relationships may be reproduced. These 
conditions may then be used as additional constraints on GC formation.

\section{Data Compilation}

The blue tilt is parameterized as $Z \sim M^{\alpha}$, where $Z$ is the GC metallicity, $M$ is the GC 
mass, and $\alpha$ is the slope of the tilt. Given their small spread in metallicity, the blue GCs are 
assumed to have a constant $M/L$ ratio. $Z$ must be calculated directly from the GC colors using a 
color-metallicity conversion. Unfortunately, nearly all of the individual studies have utilized different 
filters, so different (and possibly inconsistent) color-metallicity conversions must be used. There is no 
compelling evidence that a more complicated function than a single power law is favored.

The basic method is the same in all papers. The color-magnitude diagram is divided into magnitude bins, 
and within each bin Gaussian mixture modeling is performed to fit the mean color of each subpopulation 
within the bin. The slope estimates come from linear fits to the resulting mean colors of each 
subpopulation as a function of magnitude. The routine typically used for the mixture modeling is KMM 
(Ashman \etal~1994), and as discussed in Mieske \etal~(2006), KMM can introduce a systematic error into 
the derived peak locations if there is a large imbalance in the number of objects between the peaks in the 
usually bimodal color distribution. This can occur if the magnitude binning is done in a blue filter (for 
example, $B$ or $g$), since the large metallicity differences between the subpopulations lead to different 
cluster mass-to-light ratios at blue wavelengths. By contrast, red filters (e.g., $R$, $I$, or $z$) show 
little or no $M/L$ variations in the relevant metallicity range. All of the results cited below utilize a 
red filter for binning and so should not suffer from this systematic error. We note that mixture modeling 
with Nmix (Richardson \& Green 1997), as in Strader \etal~(2006) and Spitler \etal~(2006), is more robust 
to this effect than KMM. Observations of Galactic GCs (e.g., Pryor \& Meylan 1993; McLaughlin \& van der 
Marel 2005) also show little or no correlation between metallicity and mass-to-light ratio.

Harris \etal~(2006) studied the blue tilt in eight massive Es that are at the center of their respective 
group or cluster. The slope of the composite blue tilt is $\alpha = 0.55\pm0.06$, and the blue tilt 
appears at $M_I \la -9.5$, corresponding roughly to cluster masses of $M > 6 \times 10^5 M_{\odot}$. Below 
this ``break" magnitude there is no evidence for a variation in the mean color. It is worth noting the 
diversity of the GC luminosity at which the tilt appears in their sample---for NGC 3268 and NGC 3258 it is 
visible at $M_I \sim -8.5$ (near the GC luminosity function turnover at $2-3 \times 10^5 M_{\odot}$), 
while in NGC 4696 the tilt does not become obvious until $M_I \la -10$ (nearly $10^6 M_{\odot}$). The 
photometry is relatively deep and these differences appear genuine. Due to the merging of the blue and red 
peaks at the brightest magnitudes, and the small number of very luminous GCs present, the precise behavior 
of the blue tilt at the bright end is unclear. Harris \etal~(2006) estimate metallicities using the 
relation [Fe/H] = $2.667 (B-I) - 5.757$, derived from a fit to Galactic GCs. Using this relation, the mean 
metallicity at the break mass of the blue tilt is [Fe/H] $\sim -1.4$, increasing to [Fe/H] $\sim -1.0$ for 
the most luminous GCs. To illustrate the systematic uncertainties in the color-metallicity conversion, 
Smith \& Strader (2007) find [Fe/H] = $1.898 (B-I) - 4.748$ from Galactic GCs with low reddening and 
[Fe/H] $< -0.5$; using this relation instead would give a slope $\alpha = 0.39\pm0.04$. Since there is 
some evidence for nonlinearity in color-metallicity relationships (e.g., Strader \etal~2007 and references 
therein), the latter slope may be more accurate, but this is uncertain at present.

More recently, using Gemini/GMOS imaging of NGC 3311 (the massive central galaxy in the Hydra Cluster), 
Wehner \etal~(2008) found the blue tilt in $g-i$. The estimated slope is $\alpha = 0.6\pm0.2$, using a 
conversion between $g-i$ and metallicity derived from stellar population models: $(g-i) \propto 0.18$ 
[m/H]. This paper is notable in demonstrating the blue tilt using ground-based imaging; the tilt is not a 
result of any artifact from Hubble Space Telescope observations (as claimed by Kundu 2008). The break mass 
is poorly determined but the tilt appears to extend to at least the GCLF turnover.

Spitler \etal~(2006) find a slope of $\alpha = 0.29\pm0.04$ in NGC 4594 using $B-R$ photometry. In NGC 
4594 the blue tilt extends at least to the GCLF turnover (at $\sim 2 \times 10^5 M_{\odot}$) and 
perhaps fainter. The exact magnitude at which the tilt begins is uncertain due to the larger 
photometric errors at these magnitudes. The color-metallicity relation used is [Fe/H] = $3.06 (B-R) - 
4.90$, derived from Galactic GCs with low reddening and a few NGC 4594 GCs with spectroscopic 
metallicities at the metal-rich end.

Using $g-z$ colors, Strader \etal~(2006) detected the blue tilt in two massive Virgo Es (M87 and 
M60) with a firm non-detection in M49. These same galaxies were studied in the larger sample of Virgo 
cluster galaxies in Mieske \etal~(2006). Since the measured blue tilt slopes (in $g-z$) for M87 and 
M60 agree to within the errors for the two papers, we adopt the Mieske \etal~(2006) values for 
consistency. They used the relation [Fe/H] = $5.14 (g-z) - 6.21$ (from Peng \etal~2006), derived from 
Galactic GCs with low reddening and some M49 and M87 GCs with spectroscopic metallicities. This fit 
excludes objects with high metallicity. The resulting slopes for M87 and M60 are $\alpha = 
0.54\pm0.19$ and $0.36\pm0.11$. For these massive Es, the blue tilt clearly extends to at least the 
GCLF turnover, and perhaps even a magnitude fainter, near $10^5 M_{\odot}$.

Mieske \etal~(2006) also calculate composite blue tilts for all of their sample galaxies in four bins of 
host galaxy luminosity. The first three bins all show evidence for the blue tilt, with slopes from $\alpha 
= 0.48$ to 0.41. In the faintest bin (galaxies with $M_B > -18.4$) the blue tilt is detected at only 
$1\sigma$. We stress that the slope values given are derived from composite fits, and the actual tilt 
slopes may vary substantially among galaxies.

It is unknown how many galaxies besides M49 do not have a blue tilt. Figure 2 shows GC mass vs.~[Fe/H] for 
Galactic GCs using data from Harris (1996; updated in 2003), overplotted with a fiducial blue tilt with 
$\alpha = 0.4$ that starts at the GCLF turnover. There is no evidence for the blue tilt in the Milky Way, 
though current catalogs of photometry and metallicity for Galactic GCs are inhomogeneous, and the Milky 
Way has many fewer clusters than in the giant Es in which the blue tilt has been typically detected. 
Figure 2 also shows that if the blue tilt were only present for masses $\ga 10^6 M_{\odot}$, it would be 
undetectable.

To summarize, the range of well-measured slopes is $\alpha \sim 0.3-0.6$, with a ``typical" slope of $\sim 
0.4$. The GC mass at which the tilt begins is not always well-determined, but appears to range from at 
least $2-10 \times 10^5 M_{\odot}$. In the following sections we draw on similar discussions in Dekel \& 
Silk (1986), Dekel \& Woo (2003), and Ashman \& Zepf (2001).

\section{Models Without Dark Matter}

\subsection{The Cluster Mass-Metallicity Relation}

The simplest case we can consider is GC formation from an isolated, self-gravitating gas cloud with no 
mass loss. The mean metallicity $Z_m$ is determined by the simple closed-box model (e.g., Tinsley 1980), 
in which case $Z_m$ is equal to the assumed yield. There is no dependence on cluster mass, and thus 
$\alpha = 0$. In fact, this closed-box model would not predict any metallicity spread among GCs unless 
there was a range in the stellar yields.

We next consider the case, still assuming a self-gravitating cloud, in which star formation stops when the 
energy from supernovae is comparable to the initial binding energy of the gas cloud. Since the former is 
assumed proportional to the mass of stars formed ($M_s$), we have

\begin{equation} 
M_s \propto M_g^2/R 
\end{equation}

\noindent
at the cessation of star formation, where $M_g$ is the initial gas mass in the cloud, and $R$ is the 
radius. This equation assumes that the mass of the protocluster cloud is still dominated by gas at the 
time that star formation is turned off (i.e., $M_s << M_g$)\footnote{Relaxing this assumption makes the
model only solvable numerically in most cases and obscures the relationship between the input parameters 
and results.}, so 
that the stellar mass contributes negligibly to 
the binding energy of the gas. We 
emphasize that this assumption applies globally to the cloud, not locally in the star-forming core (see 
discussion below). We then assume a mass-radius relationship for the clouds of the form

\begin{equation}
M_g \propto R^n.
\end{equation}

\noindent
A power-law index $n=2$ corresponds to clouds in virial equilibrium, and $n=3$ to equal density 
clouds. Combining relationships gives

\begin{equation}
M_s \propto M^{2-(1/n)}_g.
\end{equation}

The mean metallicity $Z_m$ of the cloud and its subsequent GC after self-enrichment scales with 
$M_s/M_g$ (assuming zero initial metallicity), which leads to

\noindent
\begin{equation}
Z_m \propto M_s^{(n-1)/(2n-1)},
\end{equation}

so the final equation for $\alpha$ is: $\alpha=(n-1)/(2n-1)$. This gives $Z_m \sim M_s^{1/3}$ for $n=2$ and $Z_m \sim 
M_s^{2/5}$ for $n=3$, in the general range of observed values of $\alpha$ (though see caveats in \S 5). 
The asymptotic 
behavior for large $n$ is $\alpha = 0.5$ (this corresponds to the case in which all clouds have the same radius); 
steeper slopes are not possible. If $n=1$ there is no blue tilt.

For this model we have not specified any value for the efficiency factor of GC formation $\epsilon = 
M_s/M_g$ (that is, the amount of mass in the protocluster cloud that ends up in the bound star 
cluster), although the above scenario implies that $\epsilon$ should scale with $M_s$ in an identical 
manner to the metallicity $Z_m$, i.e.,

\begin{equation}
\epsilon \propto M_s^{(n-1)/(2n-1)}.
\end{equation}

By neglecting the contribution of the star cluster to the binding energy of the protocloud gas, we have 
assumed that the mass of the cluster formed $M_s$ is significantly smaller than the mass $M_g$ of the 
cluster protocloud, so the global efficiency $\epsilon$ is small. By contrast, it is generally thought 
that for an individual star-forming cloud core, the local efficiency must be high ($\sim 0.3$ or higher) 
for the resulting star cluster to remain bound (e.g., Boily \& Kroupa 2003). Thus our assumptions 
correspond to a scenario in which the bulk of a star cluster forms from a small core within a larger 
parent cloud. Within this core the star formation efficiency is high, while averaged over the entire 
cloud it is much lower. This is in accord with the observations of molecular clouds and young star 
clusters summarized by Ashman \& Zepf (2001). Energy from cluster supernovae becomes dispersed 
throughout the parent cloud and eventually causes star formation to cease, and possibly the dispersal of 
the parent cloud. We note again here that we are assuming a single generation of star formation, and 
expect that self-enrichment will lead to metallicity inhomogeneity in individual GCs.

It would be possible to generalise our scenario to a two-generation model in which a first generation of 
supernove enrich a second generation of stars that forms at some later time, after the SN ejecta have been 
homogeneized into the protocluster gas cloud. This would involve relating the star formation efficiency of 
the first stellar generation to the second, and so would introduce more free parameters into the scenario 
without necessarily producing a chemically homogeneous cluster.

\subsection{The Cluster Mass-Radius Relation}    

The scaling arguments above have not considered the radius $R_s$ of the star cluster that eventually 
results after gas loss from the parent cloud. We obtain a scaling relationship for $R_s$ by further 
assuming that the GC forms from a core within a much larger parent cloud in which there is an $r^{-2}$ 
density distribution, such that the mass within radius $r$ is $M(r) \propto r$. The star cluster forms 
from a central core of initial radius $R_c$ and mass $M_c$ within this cloud. The star formation 
efficiency in the core is $\epsilon_c = M_s/M_c$, which is distinct from the efficiency 
$\epsilon$ averaged over the entire parent cloud of mass $M_g$. The assumed density distribution 
within the cloud leads to $M_c/M_g = R_c/R$, so that $R_c = (\epsilon/\epsilon_c) R$. With relations 
(2) and (5) we have

\begin{equation}  
R_c \propto  (1/\epsilon_c) M_s^{n/(2n-1)}.
\end{equation}

Suppose that the core expands adiabatically as gas is lost and the core mass drops from the original 
value of $M_c$ within a radius $R_c$ to the final mass of $M_s$ and radius $R_s$. If gas is lost slowly 
compared to a dynamical time then $M_s R_s = M_c R_c$ (e.g, Hills 1980, Ashman \& Zepf 2001), which 
leads to

\begin{equation}
R_s \propto  (1/\epsilon_c)^2 M_s^{n/(2n-1)} \propto  (1/\epsilon_c)^2 M_g
\end{equation}

\noindent
If $\epsilon_c$ within the protocluster core scales as $M_g^{1/2}$ then there is no cluster 
mass-radius relation. For a constant $\epsilon_c$, for all $n$ it is not possible to erase
a GC mass-radius relation.

If we set $\epsilon = \epsilon_c$ such that $R_c = R$, i.e., the cluster forms throughout the entire 
cloud (although this seems less compatible with our assumption of $M_s << M_g$), then upon using 
relations (5) and (7) we get 

\begin{equation}
R_s \propto M_s^{(2-n)/(2n-1)}.
\end{equation}

\noindent
Thus in the case where a star cluster forms throughout the full volume of the parent cloud, a mass-radius 
relationship will be erased if $n=2$. This is analogous to the situation considered by Ashman \& Zepf (2001), who by 
assuming initially virialized gas clouds, showed that the GC mass-radius relation can be eliminated if it is assumed 
that $\epsilon \propto M_g/R$, which in our notation reduces to $M_s \propto M_g^2/R$ (our equation 1). The point 
here is that, rather than an arbitrary solution to the lack of a mass-radius relation for GCs, the Ashman \& Zepf 
(2001) condition naturally emerges from the hypothesis that supernova feedback governs star formation in massive 
clusters. By contrast, Ashman \& Zepf (2001) note that $M_g/R$ is the binding energy per unit mass, and suggest that 
a different interpretation of the $\epsilon \propto M_g/R$ condition is that more tightly bound proto-GC clouds can 
more efficiently convert gas into stars. As shown above, such an assumption will lead to a blue-tilt relation 
between $Z_m$ and $M_s$ for the resulting GCs.

In our scenario of a GC forming within the central core of a much larger cloud, if we 
set $\epsilon_c \propto \epsilon$ then we can again retrieve a cluster mass-radius relation of $R_s 
\propto M_s^{(2-n)/(2n-1)}$. Thus although 
$\epsilon_c$ is much larger than $\epsilon$, if they scale together as a function of cloud mass then 
we again can erase the cluster initial mass-radius relation for $n=2$.

A small modification of the above picture is to assume that the individual clouds do not simply follow a power 
law mass-radius relation but instead follow the specific Bonner-Ebert relation for pressure-confined 
clouds (as do giant molecular clouds in the Galaxy). Using the relation $R \propto M_g^{1/2} P^{-1/4}$ 
(Ashman \& Zepf 2001), where $P$ is the surface pressure of the cloud, gives a final mass-metallicity 
relation $Z_m \sim M_s^{1/3} P^{1/6}$. The dependence on the pressure is weak but could lead to some 
modification of the cluster metallicities in different environments.

In the above discussion we have neglected the contributions of any pre-enrichment to the metal content 
of a GC protocloud. While this might be approximately true for blue GCs, it is unlikely 
to be the case for the red GC subpopulation of a galaxy. In fact, a self-enrichment origin for the 
blue tilt could be consistent with the lack of a metallicity-mass relation among the red 
subpopulations of GCs within galaxies. Suppose that the initial metallicity of a blue GC protocloud is 
$-1.5$ dex, equivalent to an initial $Z_{i} = 6 \times 10^{-4}$. This is the mean pre-enrichment 
metallicity of a blue GC. Further suppose that self-enrichment leads to a metallicity for the final GC 
that is 0.4 dex higher (a typical value for the brightest blue GCs), by adding an extra $1.5 Z_i$ 
metals per unit mass. In the case of a red GC the initial pre-enrichment metallicity might be as much 
as $10 Z_i$. In this case, adding another $1.5 Z_i$ of metals per unit mass (assuming that the amount 
of enrichment does not scale as $Z_i$) gives an enhancement in metallicity of 0.06 dex. It is unlikely 
that such an effect could be readily detectable as a ``red tilt.''

\subsection{Globular Cluster and Cloud Mass Functions}

An observable implication of this model is that the mass function of the GCs is, in most cases, flatter than that 
of the clouds themselves. This conclusion was reached by Ashman \& Zepf (2001), whose results we recover as a 
specific case below.

First we consider the case where star formation is dispersed throughout the cloud. Assume that the 
protocluster clouds and GCs have power law mass functions over the mass range of relevance, with slopes 
$\beta_{i}$ and $\beta_{f}$, respectively. We then have:

\begin{equation}
M_s^{-\beta_{f}} dM_{s} \propto M_g^{-\beta_{i}} \frac{dM_{g}}{dM_{s}} dM_{s}.
\end{equation}

\noindent
Using (3) and equating the exponents, we derive the general equation for $\beta_{f}$

\begin{equation}
\beta_{f} = \frac{n(\beta_{i}+1)-1}{2n-1}
\end{equation}

\noindent
For virialized clouds with $n=2$, $\beta_{f} = 2(\beta_{i}+1)/3$, recovering the result of Ashman \& Zepf (2001). 
$\beta_{f} < \beta_{i}$ for all $n > 1$, and for large $n$ the equation approaches $\beta_{f} = (\beta_{i}+1)/2$.

Rearranging the equation $\alpha = (n-1)/(2n-1)$, we can derive the predicted slope of the blue tilt $\alpha$ as 
a function of $\beta_{i}$ and $\beta_{f}$

\begin{equation}
\alpha = (\beta_{i} - \beta_{f})/(\beta_{i} - 1)
\end{equation}
 
This equation is remarkable in that it relates the predicted blue tilt slope $\alpha$ to the 
quantities $\beta_{i}$ and $\beta_{f}$. For the case $\beta_{i} = 2$ this equation reduces to the 
admirably simple $\alpha = 2 - \beta_{f}$.

Next we consider the case governed by equation (7) in which the GC forms in the core of a star-forming cloud. 
We assume the cloud core mass function follows a power law with slope $\gamma_{i}$. To make the problem 
tractable, we also assume the condition necessary to erase a GC mass-radius relationship, 
$\epsilon_c \propto M_{g}^{1/2}$. Combining this condition with equation (3) gives an equation analogous to (10):

\begin{equation}
\beta_{f} = \frac{n}{2(2n-1)} + \gamma_{i}[1-\frac{n}{2(2n-1)}]
\end{equation}

\noindent
Though (12) looks different than (10), for the special case $n=2$, it reduces to the equivalent result $\beta_{f} 
= 2(\gamma_{i}+1)/3$. This is due to the self-similar nature of the cloud. For other $n$, the cloud and core 
results are generally similar but not identical. For example, if $n=3$ in (10), then for $\beta_{i} = 2$, 
$\beta_{f} = 1.6$. In the core case, for $n=3$ and $\gamma_{i} = 2$, $\beta_{f} = 1.7$.

Replacing $n$ with $\alpha$ gives

\begin{equation}
\alpha = \frac{2\beta_{f}-(\gamma_{i}+1)}{\gamma_{i}-1}.
\end{equation}

\noindent
At fixed $\alpha$, the difference between the cloud or core mass function slopes $\beta_{i}$ and $\gamma_{i}$ and 
the GC slope $\beta_{f}$ is smaller in the core case for the observed range of $\alpha$, $0.3 \la \alpha \la 
0.5$. Recall that in this model $\alpha > 0.5$ cannot be produced.

One can imagine two different applications of these equations. For observed cloud (or core) and GC mass function 
slopes, $\alpha$ can be predicted. Alternatively, if one assumes that our self-enrichment model is correct, then one 
can use the observed values of $\alpha$ and $\beta_{f}$ in a galaxy to deduce the initial values of $\gamma_{i}$ or 
$\beta_{i}$.

\section{Dark Matter Model}

Here we investigate the case in which GCs form in individual dark matter halos---that is, they are 
``mini-galaxies". In a proper cosmological setting, the clouds considered in \S 3 would also be hosted 
by dark matter halos. The crucial distinction is between a scenario in which the clouds are 
self-gravitating and one in which the potential well of the host halo dominates.
	
The basic model for the evolution of a GC in a dark matter halo is similar to that discussed in \S 3. 
Here we follow the derivation in Dekel \& Woo (2003). The mass of the proto-GC cloud is much smaller 
than that of the dark matter halo, and thus the formation of the cluster has essentially no effect on 
the halo. Star formation halts when $M_s \propto M_{vir} V_{vir}^2$, where $M$ and $V$ are the virial 
mass and velocity of the halo, respectively. These two quantities are related by $M_{vir} \propto 
a^{3/2} V_{vir}^3$ with $a = (1+z)^{-1}$ the expansion factor at virialization. Substitution yields $Z 
\sim M^{2/5} a^{-3/5}$. For low-mass halos the range in $a$ is expected to be small, so one simply 
reproduces the well-known result that $Z \propto M^{2/5}$. This is a good match to dwarf galaxies in 
the local universe, suggesting that their properties are consistent with supernova feedback in 
dominant dark halos. The slope is also consistent with a typical value for the blue tilt.

However, we must consider the other scaling relations expected in this model. Assuming a constant $a$, 
the radius of the cluster is given by $R_s \propto \lambda M^{1/5}_{s}$, where $\lambda$ is the 
dimensionless spin parameter of the halo. We would then require $\lambda \propto M^{-1/5}_{s}$ to 
erase the mass-radius relationship (this same point is noted in Cen 2001). More crucial is the 
absolute value of $\lambda$; halos with typical values of $\lambda \sim 0.02-0.03$ have too much 
angular momentum to collapse to anything approaching the size of a GC. Even if one assumed that GCs 
came preferably from the low-$\lambda$ tail of the spin distribution---and the values would have to be 
quite small, naively a factor of $\sim 50-100$ lower than low-mass dwarf galaxies---one would still 
expect to see a much larger population of objects with sizes between those of GCs and dwarf galaxies 
that came from ``intermediate" low-spin halos.

We can move a step further back and assume instead that we circumvent the angular momentum problem by 
forming the GC as the baryonic core of a more massive protogalaxy. The GC would be analogous to the 
nuclei of dwarf galaxies in the local universe. Such a scenario has been invoked for GCs such as 
$\omega$ Centauri (e.g., Zinnecker \etal~1988; Bekki \& Freeman 2003) and more generally by B{\"o}ker 
(2008). An upper limit to the expected effect of dark matter can be found by assuming a cusped NFW 
profile for the dark matter; actual dwarfs may instead have cored dark matter profiles (e.g., Burkert 
1995). NFW models of Milky Way dSphs, corresponding to virial masses of $\ga 10^8 M_{\odot}$, have dark 
matter masses within the central $\sim 10$ pc of $\sim 10^4 M_{\odot}$ or slightly higher (e.g., Walker 
\etal~2007). This suggests that the dark matter would have little dynamical effect on a nuclear GC of 
mass $\sim 10^6 M_{\odot}$, and that the evolution would be similar to the self-gravitating cloud model 
of \S 3.

However, there are other issues with the stripped nuclei scenario. It is uncertain whether the halo and 
other associated baryonic material can effectively be stripped for hundreds of objects in a given 
galaxy, and whether the expected structural parameters of these nuclei can match that of the GCs. For 
example, C{\^o}t{\'e} \etal~(2006) find that the luminosity function of nuclei in Virgo cluster 
galaxies peaks $\sim 3.5$ mag brighter than the GCs; the nuclei also show a steep mass-radius 
relationship. Such observations are seemingly not in accord with the GC--dwarf nuclei scenario, but 
perhaps biasing arguments could help address these issues.

\section{Subsequent Evolution of the Blue Tilt}

The models described in \S 3 and \S 4 strictly apply only to an \emph{initial} mass-metallicity 
relationship for the metal-poor GCs. Subsequent mass loss from the GCs will modify the relation. This in 
fact would be the case for any scenario in which the blue tilt is a primordial phenomenon, whether caused 
by the internal self-enrichment of proto-GCs or by some other process related to the cluster environment 
at early times.

There are three principal sources of mass loss from star clusters: (i) prompt early gas expulsion, with 
an accompanying removal of stars, (ii) mass loss due to stellar evolution, as stars eject gas via 
stellar winds and planetary nebula shells, and (iii) slow mass loss from two-body relaxation and 
subsequent evaporation that lead to the loss of individual stars from a cluster. We consider these in 
turn.

The removal of gas from nascent GCs due to supernovae weakens the cluster potential and unbinds a 
fraction of the stellar mass of the cluster. How much mass is lost depends both on the timescale of the 
gas removal (compared to the dynamical timescale of the GC) and on the efficiency of star formation.  
Numerical simulations indicate that high efficiency ($\epsilon_{c}$ in \S 3) and slow gas removal lead 
to little mass loss, consistent with our assumptions in \S 3.1 and \S 3.2 (Baumgardt \& Kroupa 2007; 
Parmentier \etal~2008). For high enough efficiencies, (say $\epsilon_c \ga 0.6-0.7$), the timescale of 
the gas removal becomes less important. This is especially true if the tidal field of the GC is weak, 
as one might expect for metal-poor GCs forming in the proto-halos of their host galaxies.

We conclude that prompt mass loss from gas expulsion does not substantially alter the slope of the blue 
tilt under the assumptions of our model. However, for lower star formation efficiencies or for very 
fast gas expulsion, our assumptions do not hold and the masses of the clusters may be changed 
substantially.  This could act to induce scatter in the blue tilt, or erase it altogether. Therefore 
the existence of a blue tilt may favor, or even require, high star formation efficiency in the core of 
a proto-GC cloud.

The second source of mass loss is stellar evolution. All stars lose mass through winds, and more massive 
stars generally have higher mass loss rates. Models of the evolution of single stellar populations suggest 
that this form of mass loss is a constant fraction of the cluster mass for a given initial mass function 
(IMF)---see, e.g., the implementation in the models of Fall \& Zhang (2001). Thus mass loss from stellar 
evolution will change only the zero-point of the blue tilt, but not its slope, as long as the IMF of 
clusters does not vary. If we assumed that the IMF was a strong function of metallicity, then this could 
have a small effect on the slope of the blue tilt.

We finally consider mass loss from the slow dynamical evolution of star clusters. Two-body relaxation 
leads to energy equipartition and a quasi-Maxwellian distribution of velocities, and the stars in the tail 
of the distribution evaporate into the field. A number of papers have investigated this form of mass loss, 
from both theoretical and observational standpoints. In particular, Jord{\'a}n \etal~(2007) use data from 
the ACS Virgo Cluster Survey to argue that the GC mass functions of early-type galaxies in Virgo are best 
fit by a model in which the amount of mass lost is independent of the initial mass of the GC, and is 
approximately equal to the classic ``turnover" at $\sim 2 \times 10^{5} M_{\odot}$.

There is no closed-form general expression to relate an initial blue tilt slope to that after a 
constant amount of mass loss from all clusters. In particular, the power-law form is not retained, 
although the deviations from a power-law are small for typical parameters and would not be detectable 
in current data. A simple example demonstrates the basic features of the model. A typical observed blue 
tilt extends from $\sim 2 \times 10^{5} M_{\odot}$ to $10^{6} M_{\odot}$ with $\alpha = 0.4$. Adding in 
a constant mass loss of $2 \times 10^{5} M_{\odot}$ and refitting over the entire mass range gives an 
inital slope of $\sim 0.59$. 
Thus constant mass loss leads to shallower observed slopes. Less mass loss naturally produces a smaller 
change; for example, mass loss of $10^{5} M_{\odot}$ would yield an initial slope of 0.50. Constant 
mass loss also has a smaller effect if only massive clusters are considered; for example, in some of 
the giant ellipticals in Harris \etal~(2006), the break mass of the blue tilt is $\sim 10^{6} 
M_{\odot}$. For these galaxies constant mass loss of $2 \times 10^{5} M_{\odot}$ would produce little 
change in the blue tilt slope.

Overall, we conclude that mass loss subsequent to GC formation is likely to change both the slope and 
zero-point of the observed blue tilt from its initial value. These changes should be taken into account 
in attempts to derive initial star formation conditions from present-day blue tilts.

\section{Blue Tilt Variations}

As described in \S 2, there are observed variations in both the slope of the blue tilt and the GC mass at which it 
becomes evident. While some of these variations may be due to modifications of GC masses subsequent to cluster formation 
(as discussed in \S 5), might there also be sources of intrinsic variation in any of the models discussed?

Within the context of a self-enrichment scenario, Recchi \& Danziger (2005) found that dispersion in a GC 
mass-metallicity relation could be attributed to variations in the thermalization and mixing efficiencies 
of supernova ejecta, the stellar initial mass function, and external protocloud pressure. In the context 
of the scaling relations presented above, the slope of the blue tilt in the self-enrichment, 
self-gravitating model depends upon the initial mass-radius relation of the clouds. If the initial 
mass-radius relationship for clouds varied among galaxies, this would lead to differences in the blue tilt 
slope. A relation between the surface pressure of the cloud and its mass might also change the expected 
tilt, although the dependence on pressure is weak. In addition, any scatter in the mass-radius 
relationship among GC parent clouds would act to erase a blue tilt among the resultant clusters. Thus the 
presence or absence of a blue tilt may reflect how well the parent clouds in a galaxy followed a 
well-behaved mass-radius law. In addition, the absence of a blue tilt in some galaxies could also be 
attributed to a protocloud mass-radius relation with $n=1$.

The blue GC populations of present-day galaxies are expected to have assembled hierarchically through 
mergers of less massive galaxies. We may then speculate that an additional factor in blue tilt 
variations among galaxies is the respective merging histories of the galaxies. If metal-poor GCs 
formed in low-mass halos at relatively high redshift (e.g., Cen 2001; Bromm \& Clarke 2002; Rhode 
\etal~2005; Brodie \& Strader 2006), then the blue tilt is expected to have been in place since then. 
The subsequent merging of halos with similar blue tilts would then produce a galaxy with that tilt; 
the merging of halos with different blue tilts (due, for example, to differing cloud mass-radius 
relationships as discussed above) could weaken or erase the progenitor tilts.

Nearly all of the blue tilts studied thus far appear to have a GC break mass at which the tilt appears. 
This mass varies among galaxies. We have assumed that GC self-enrichment is controlled by an energy 
balance between supernova feedback and the binding energy of the clouds. It is this balance which 
determines the ratio that results between parent cloud mass and the mass of cluster stars ultimately 
formed, which in turn sets the level of metallicity enrichment in the protocloud. This energy balance 
could be changed in several ways.  For example, the fraction of supernovae energy that goes into cloud 
disruption might vary with cloud density. This could occur if the ratio between the radiative timescale 
of supernovae and the dynamical time of the cloud varied substantially with the cloud mass. If the 
radiative timescale had a strong inverse dependence on cloud density, then less dense clouds 
(equivalent to more massive clouds in virial equilibrium) might require more energy per unit gas mass 
to halt star formation.  However, Dekel \& Silk (1986) argue that this ratio is approximately constant 
for the relevant range of parameters, and it is not clear why it would vary among galaxies.

Another way to affect the energy balance is if the supernova energy in a single cloud does not 
dominate the overall energy input---for example, if the clouds are not isolated, but have energy input 
from nearby star-forming regions. Consider a situation in which all of the proto-GC clouds are bathed 
with a field of uniform energy density. In the high-mass clouds this might be only a fraction of the 
energy from their own supernovae, and thus would have little effect. However, in clouds of lower mass, 
this additional energy input could serve to disrupt star formation before significant self-enrichment 
had occurred. For such clouds, the metallicity would then be set by the starting metallicity of the 
cloud. The initial cloud metallicity, set by pre-enrichment, should be independent of cloud mass. In 
this (speculative) scenario, the explanation for different blue tilt break masses would be variations 
among galaxies in external energy sources during GC formation. On the other hand, it is not clear that 
the proposed mechanism for erasing the GC mass-radius relation would still be relevant if external 
energy input was important during cluster formation.

For the dark matter scenario, there seem to be fewer natural avenues for variations in the slope of the 
blue tilt, so if further studies strengthen the case for substantial differences in slope among 
galaxies, this could be evidence against this model. On the other hand, our simple suggestion for 
variations in the break mass due to ambient radiation would naively seem to work equally well in the 
dark matter case.

\section{The Blue Tilt and Galaxy--Globular Cluster Correlations}

In the last few years it has become clear that there are correlations between the mean color (and thus 
metallicity) of both old metal-poor and metal-rich GCs and the stellar mass of their parent galaxies 
(Larsen \etal~2001; Forbes \& Forte 2001; Strader \etal~2004; Lotz \etal~2004; Peng \etal~2006; 
Strader \etal~2006). Here we consider whether the blue tilt could have a significant affect on the 
relation between the mean color of the blue GC population and galaxy luminosity.

\subsection{The Blue Tilt and Mean Globular Cluster Colors}

The mean color of blue GCs is generally determined from the brighter clusters, due to incompleteness 
and larger photometric errors for fainter GCs. The existence of the blue tilt implies a dependence of 
the estimated mean color on the limiting magnitude of the study. Let us take M87 as an example. 
Assuming that the bright half of the blue GCLF is approximately lognormal, we use the parameters 
derived in Strader \etal~(2006) for the $z$ band ($\mu_{z} = -8.2$; $\sigma_{z} = 0.92$). Mieske \etal~(2006) and 
Strader \etal~(2006) find very similar slopes for $g-z$ vs.~$z$: $-0.042$ and $-0.043$, respectively. 
Integrating from the GCLF turnover to 3 mag brighter gives an overall mean $g-z$ color 0.03 mag redder 
than the value at the turnover. In a ``worst case" scenario in which the photometry extends only to 1 
mag brighter than the turnover, the difference is large, with a mean color 0.06 mag redder than the 
turnover value.

The mean color derived from very deep photometry depends on the behavior of the blue tilt at faint 
magnitudes. If we assume that there is no tilt fainter than the turnover, then deep photometry can 
effectively dilute the blue tilt. For example, if the photometry extended to 1 mag beyond the 
turnover, the mean color would be less than 0.02 mag redder than the turnover color.

By influencing the mean color that is derived for a GC system the blue tilt can conceivably contribute 
to the relation observed between mean GC color and parent galaxy luminosity. To assess whether this
effect might be significant, let us consider the specific case of the relation obtained by 
Strader \etal~(2006) between the mean color of the blue GCs and the parent galaxy luminosity. They 
found $\langle g-z \rangle \propto -0.014 M_{B}$, implying a $\langle g-z \rangle$ color difference of 
$\sim 0.07$ mag between the most luminous Es at $M_{B} \sim -21.5$ and the bulk of the dwarf 
ellipticals with $-17 \ga M_{B} \ga -16$. The largest expected effect of the blue tilt on
the relationship between $M_{B}$ and $\langle g-z \rangle$ for blue GC systems can be 
estimated by assuming that all of the massive Es have blue tilts similar to M87 and that all dwarf Es 
have no tilt. In reality, some massive Es (like NGC 4472) have no tilt, and there may be a weak tilt 
in the dwarfs (Mieske \etal~2006).

The $\langle g-z \rangle$ colors in Strader \etal~(2006) were estimated from photometry that reached 
typically 0.5--0.7 mag fainter in $z$ than the GCLF turnover. Using the model outlined above, we then 
expect the massive Es to have mean colors $\sim 0.02$ mag redder ($\sim 0.1$ dex more metal-rich) than 
the turnover color. Since the observed $\langle g-z \rangle$ color of blue GCs is $\sim 0.07$ mag 
redder in giant Es than in dwarf Es, the blue tilt does not have a dominant effect on the slope of the 
blue GC mean color--galaxy luminosity relation, even using generous assumptions.

In a realistic model, in which a fraction of all galaxies have no blue tilt, and less luminous 
galaxies have a weaker (but still detectable) tilt, then the expected change could be substantially 
less than 0.02 mag. At least for moderately deep photometry, the blue tilt appears to be 
a minor source of scatter in the blue GC color--galaxy luminosity correlation, appearing only at the 
0.01--0.02 mag level.

Even though galaxy-to-galaxy variations in the blue tilt do not appear likely to explain 
the relation between mean GC colors and parent galaxy properties in extragalactic GC systems, it is 
worth formalizing more precisely how the mean color of a GC system is to be defined. In the presence 
of a blue tilt, perhaps some fiducial magnitude range should be defined for the GCs from which the 
mean color is determined, or else the GC color at some reference magnitude such as the peak in the 
GCLF could be adopted. With sufficiently deep observations it would be valuable to determine if there 
is a correlation between parent galaxy magnitude and the mean color of those GCs that are fainter than 
the blue tilt break magnitude.

\subsection{Setting the Red Globular Cluster Metallicities}

Regardless of these pragmatic issues, there is the deeper question of whether the origin of the blue tilt relates 
to the correlations that have been identified between galaxy absolute magnitude and the mean metallicities of 
both blue and red GC subpopulations. The blue tilt, as we have interpreted it in \S 3, is a product of the 
self-enrichment of GC protoclouds. By contrast, the mean GC color--galaxy luminosity relations refer to the 
average metallicity of ensembles of clusters. Given that the red GCs do show a correlation between mean color and 
parent galaxy luminosity, but do not show a red equivalent of the blue tilt phenomenon, we would infer within the 
context of \S 3 that self-enrichment is not the underlying cause of the color--galaxy luminosity relation for red 
GCs. Rather this later relation would be attributed to the result of their parent galaxies influencing the 
pre-enrichment level of a red GC. In the case of the red GCs this pre-enrichment level swamps any self-enrichment 
component, as discussed in \S 3.

If the red GCs trace the bulk chemical evolution of the galaxy spheroid then an ``open-box model'' 
with mass loss could apply. Such models were introduced for the GC system of the Milky Way by Hartwick 
(1976) and Searle \& Zinn (1978) and have been applied more recently by VanDalfsen \& Harris (2004). 
In this case the mean metallicity of the red GCs would depend on two factors: (i) the ratio 
between the star formation rate and the rate of mass loss from the spheroid component of which the red 
GCs are members, and (ii) the gas-to-star mass ratio at the end of the era of red GC formation (which 
might have been quite high). Systematic variations of either one of these factors could produce a red 
GC color--galaxy luminosity relation.

If the red GCs form as part of an early spheroidal system that behaves like a ``closed box'' (no mass 
loss from the spheroid), and they trace the chemical evolution of that larger halo (as opposed to 
being self-enriched) then the characteristic metallicity of the red GCs is $Z_{red} \sim Z_i + y 
M_s/M_g$, where $y$ is the stellar yield, $M_s$ is the total mass within the galaxy that has been 
formed into stars at the time that red GC formation concludes, $M_g$ is the gas mass of the galaxy, 
and $Z_i$ is the initial metallicity of the spheroid. (This approximation comes from the closed-box 
relation $Z = Z_i + y$ ln $(M_{tot}/M_g)$ in the limit where $M_s << M_g$ for $M_{tot} = M_s + M_g$.) 
Thus there would be a relation between the average metallicity of red GCs and the efficiency of star 
formation during the early stages of galaxy chemical evolution wherein the red GCs are formed. In such 
a context, the observation that $Z_{red}$ correlates with the absolute magnitude of the parent galaxy 
indicates that in more massive galaxies the early star formation efficiency was greater. This argument 
gets modified if there is gas loss from the galaxy, whereupon the $Z_{red}-M_{tot}$ relation would 
imply that within the more massive galaxies mass loss has been less effective during the era of red GC 
formation. This is analogous to conclusions that mass loss driven by galactic winds can explain a 
relation between the mean metallicity of a galaxy and the present mass of that galaxy (e.g., Dekel \& 
Silk 1986; Tremonti \etal~2004; Finlator \& Dav\'{e} 2008). What is distinctive about the GC 
$Z_{red}-M_{tot}$ relation is that it is not tracing the entire chemical evolution history of the 
parent galaxy, but only the spheroidal component with which GC formation is associated.

In the case of the blue GCs, the galaxy-dependent pre-enrichment is much smaller---hence the 
detectability of a blue tilt relation among such systems (\S 3). In this circumstance, we have a 
picture in which the more massive blue GCs trace a combination of pre-enrichment and cluster 
self-enrichment, while the red GCs trace the overall early evolution of the parent spheroid component 
through their pre-enrichment. It is among the low-mass blue GCs, wherein the blue tilt is not evident 
(thus far), that we can observe the pristine pre-enrichment levels of the environments that formed the 
blue GCs.

\section{Discussion}

We have examined a number of scenarios for the blue tilt, a mass-metallicity relationship for 
metal-poor GCs observed in many external galaxies. The model that seems to best reproduce current 
observations is self-enrichment in proto-GC clouds; in this model, star formation is governed by 
supernova feedback, and the efficiency is a function of the protocloud mass. Certain efficiency scalings
produces no mass-radius relation for the resultant GC (in accordance with observations), and are equivalent to an 
energy balance condition for star formation. We have also speculated as to how the observed variations 
in the slope of the blue tilt and the break mass might be reproduced. Contrary to the comments in 
Strader \etal~(2006), a dark matter halo is not necessarily required for effective self-enrichment, at 
least given the assumptions of the current model.

Many of the color-metallicity relations used are uncertain, and may be the source of some of the 
observed scatter in blue tilt slopes. Since it is unlikely that these relations will see significant 
improvement soon, it would be useful to make photometric observations of different galaxies with a 
common color (e.g., $B-I$ or $g-z$) so that the intercomparisons are at least consistent. Another 
route is to directly estimate the metallicities of individual clusters using spectroscopy, or with 
near-IR imaging, which is more sensitive to metallicity than optical colors. Most of the break masses 
are poorly determined, so deeper imaging of selected galaxies would be beneficial.

The assumption of a variable formation efficiency can be tested directly by measuring cloud and 
cluster masses in nearby systems of young massive clusters, as suggested by Ashman \& Zepf (2001). 
With ALMA, accurate cloud masses and radii can be determined in such systems, testing both the 
efficiency of cluster formation and the initial mass-radius relationship for clouds as a function of 
external parameters such as pressure.

Self-enrichment has been separately proposed to explain a number of abundance anomalies in Galactic 
GCs, primarily involving light elements such as C, N, O, and Mg. The culprits are generally argued to 
be an unknown combination of massive main-sequence stars and intermediate-mass AGB stars. A natural 
expectation of self-enrichment models for the blue tilt would be the appearance of abundance anomalies 
associated with blue tilt GCs. The integrated abundances of C, N, and Mg can all be estimated in 
extragalactic GCs from low-resolution spectra. For example, Cenarro \etal~(2007) find CN line 
strengths for blue tilt GCs in NGC 1407 that are much larger than in Galactic GCs at comparable 
metallicities; unfortunately, less luminous GCs in NGC 1407 have not been observed. Studying the 
abundance patterns of GCs both above and below the blue tilt break mass could constrain the timing 
and duration of the hypothesized self-enrichment events.

The self-enrichment model makes another prediction that was described in \S 1 and \S 3. Unless star 
formation in the GC is oddly segregated by mass, such that the high mass stars form, explode as 
supernovae, and enrich the intracluster gas before any low-mass stars form, then the blue tilt GCs would 
be expected to have internal dispersions in not just the lighter elements such as C through Mg, but also 
in heavier elements produced by massive star supernovae. In the Milky Way the only GC known to have a 
large star-to-star spread in the $\alpha$-elements heavier than Mg is $\omega$ Cen (e.g., Freeman \& 
Rodgers 1975; Smith et al. 2000), a cluster that has been viewed as a test case of self-enrichment and 
chemical evolution (e.g., Carraro \& Lia 2000; Ikuta \& Arimoto 2000; Platais et al. 2003; Marcolini et 
al. 2007), and possibly M22 (Norris \& Freeman 1983; Lehnert et al. 1991; Anthony-Twarog et al. 1995). The 
majority of Galactic GCs are homogeneous to a high degree in the $\alpha$-elements.

However, the Milky Way GC system is thought not to exhibit a blue tilt (although see 
Parmentier \& Gilmore 2001), so there is no conflict with the self-enrichment scenario for 
this phenomenon. However, the challenge in other galaxies could be two-fold: (i) if blue tilt 
GCs are chemically homogeneous, then the self-enrichment picture may be less compelling; (ii) 
if blue tilt GCs are typically inhomogeneous in heavy elements, why is this not true in 
Galactic GCs of comparable mass? Regarding (i), it is intriguing that four of the most 
massive GCs in M31 have large internal metallicity spreads (Fuentes-Carrera \etal~2008). 
Forthcoming homogeneous photometric and spectroscopic datasets should allow the detection of 
a blue tilt in M31, if it exists. From a theoretical standpoint, self-enrichment models such 
as those of Brown \etal~(1991) attempt to reconcile supernova-induced enrichment with 
chemical homogeneity. With regard to (ii), we suggest reluctantly that it might be necessary 
to invoke fundamental differences between the parent clouds of Galactic GCs and those in blue 
tilt galaxies.

\acknowledgments

We acknowledge support by the National Science Foundation through Grants AST-0507729 and 
AST-0406988. J.~S.~was supported by NASA through a Hubble Fellowship. We thank A.~Dupree and 
L.~Chomiuk for useful comments. The final version of the paper owes a considerable debt to
the efforts of an anonymous referee.

\newpage

\begin{figure}
\plotone{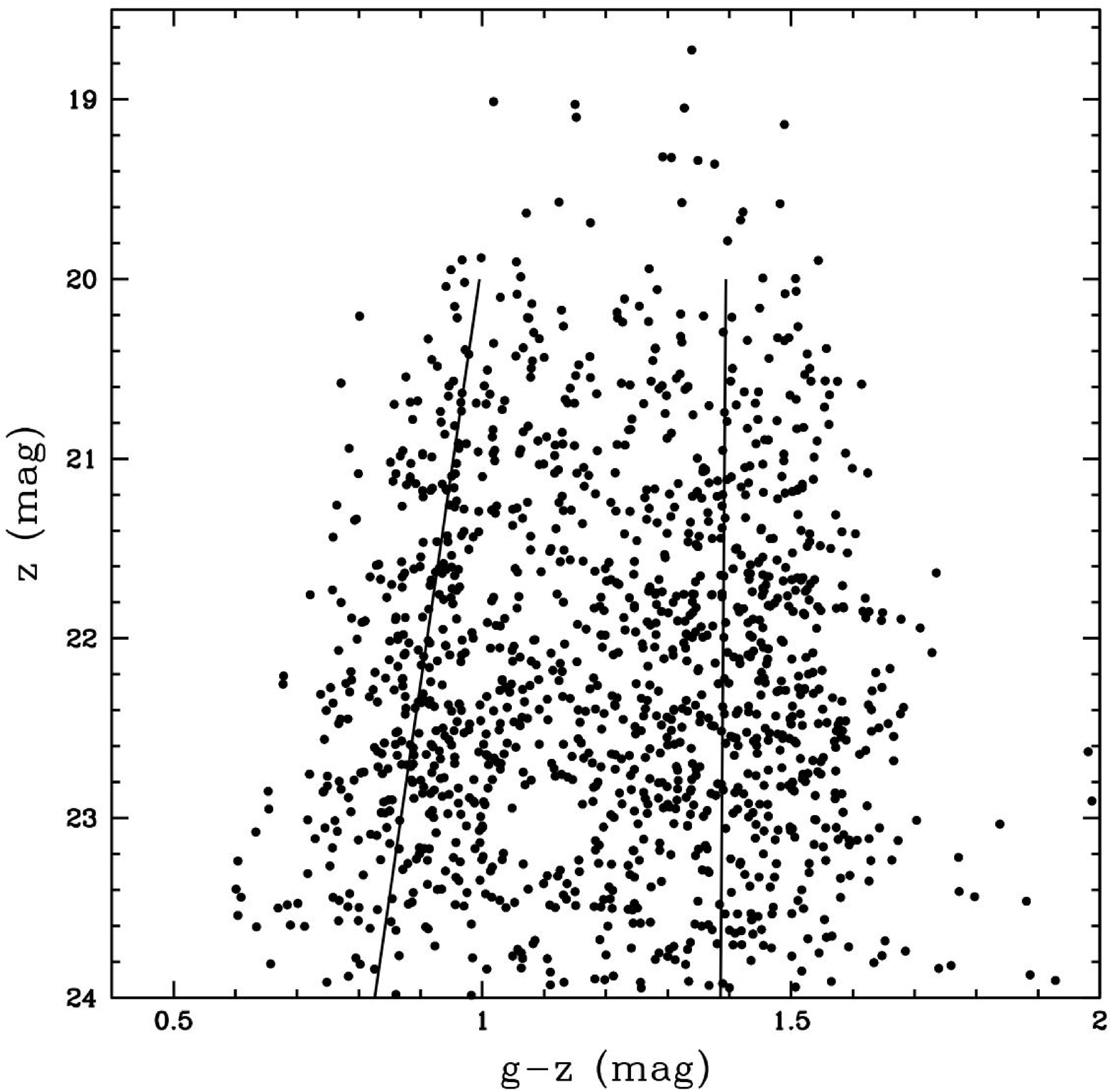}
\figcaption[f1.eps]{\label{fig:fig1}
The $z$ vs.~$g-z$ color-magnitude diagram for GCs in M87, taken
from Strader \etal~(2006). Mixture modeling of GC colors in different bins
in $z$ magnitude yields peak locations, and the overplotted linear fits to these points
show the blue tilt (and lack of a corresponding red tilt).}
\end{figure}

\begin{figure}
\plotone{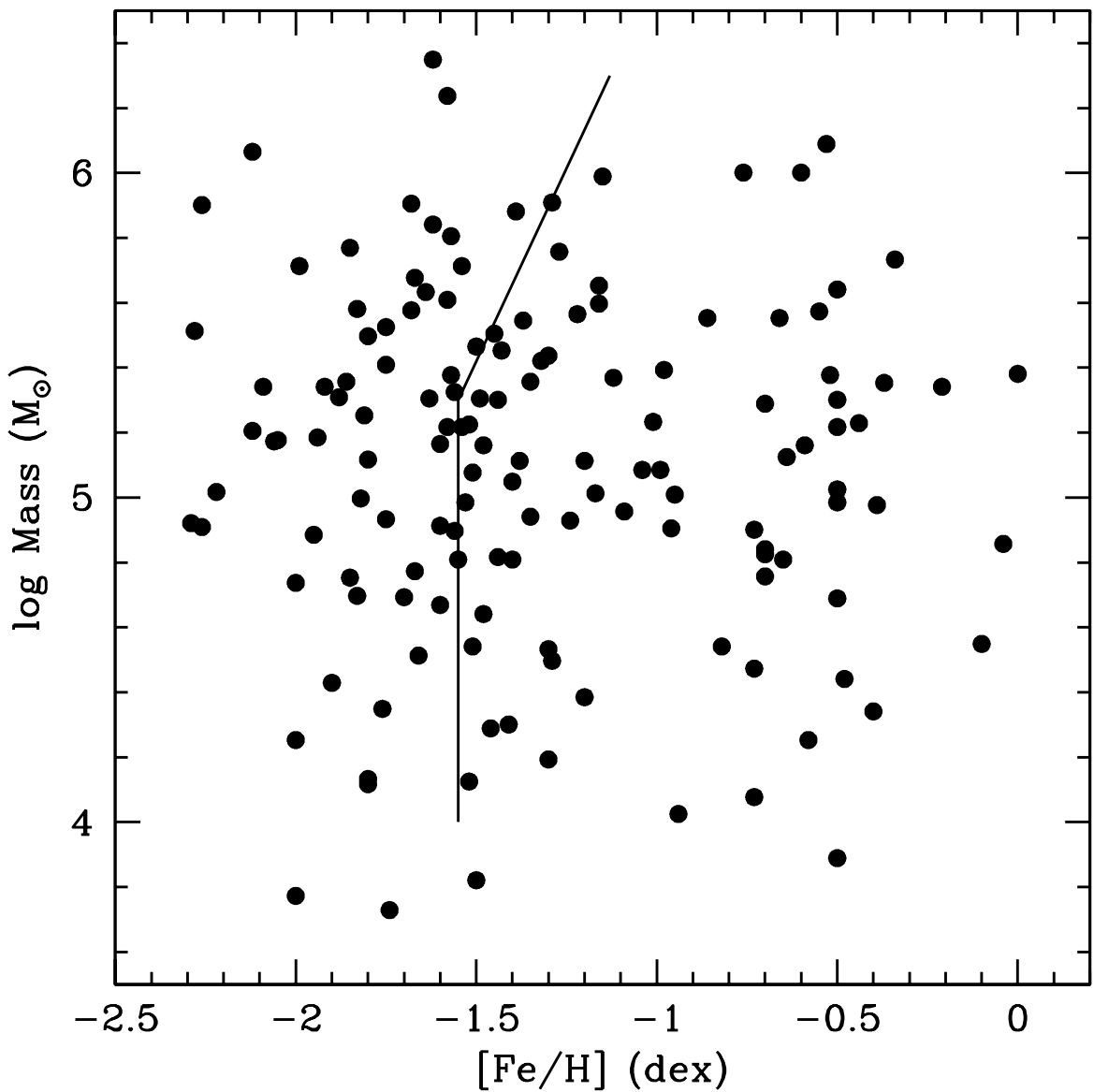}
\figcaption[f2.eps]{\label{fig:fig2}
Mass vs.~[Fe/H] for Milky Way GCs. Data is from Harris (1996), and masses have
been derived from $M_V$ assuming a constant $M/L = 2$. Though there are
systematic and random errors in the metallicities, most of the observed spread
within each metallicity subpopulation at fixed mass is real. A fiducial blue tilt
with $\alpha=0.4$, starting at a mass of $2 \times 10^5 M_{\odot}$, is overplotted.
There is no evidence that the Milky Way GCs have a blue tilt.}
\end{figure}

\end{document}